\documentclass[12pt]{article}
\usepackage{amsfonts}

\def\back{}
\def\G{\Gamma}
   
\begin{document}     

\par

\begin{flushright}
{\bf IFUM 669/FT} 
\end{flushright}
\vskip 2 truecm
{
\huge
\bf
\begin{center}
Algebraic Aspects of the\\
Background Field Method
\end{center}
}
\vspace{0.6cm} 

\begin{center}
{\Large Ruggero Ferrari\footnote{ruggero.ferrari@mi.infn.it},
 Marco Picariello\footnote{marco.picariello@mi.infn.it} and 
 Andrea Quadri\footnote{andrea.quadri@mi.infn.it}}
\\
\vskip 0.5 truecm 
Dipartimento di Fisica, Universit\`a di Milano
via Celoria 16, 20133 Milano, Italy \\
and  INFN, Sezione di Milano \\

\end{center}

\vskip 0.6 cm

\vskip 1 cm

{\bf \centerline{Abstract}}
{
\small
\begin{quotation}
The background field method allows the 
evaluation of the effective action by exploiting the (background)
gauge invariance, which in general yields Ward identities, i.e. 
linear relations among the vertex functions.
\\
In the present approach an extra gauge fixing
term is introduced right at the beginning in the action and it
is chosen in such a way that BRST invariance is preserved.
The background effective action is considered and it is
shown to satisfy both the Slavnov-Taylor (ST) identities and the Ward
identities. This allows the proof of the background
equivalence theorem with the standard techniques. In particular
we consider a BRST doublet where the background field enters
with a non-zero BRST transformation.
\\
The rational behind the introduction of an extra gauge fixing term is 
that of removing the singularity of the Legendre transform
of the background effective action, thus allowing the construction
of the connected amplitudes generating functional $W_{\rm bg}$.
By using the relevant ST identities
we show that the functional $W_{\rm bg}$ gives the same
physical amplitudes as the original one we started with.
Moreover we show that $W_{\rm bg}$ cannot in general 
be derived from a classical action by the Gell-Mann-Low
formula.
\\
As a final point of the paper we show that the BRST doublet generated
from the background field does not modify the anomaly
of the original underlying gauge theory. The proof is algebraic
and makes no use of arguments based on power-counting.
\end{quotation}

}

\eject

\section{Introduction}
\label{sec:intro}
The background field method \cite{the_beginning,abbott1,abbott2} allows the 
evaluation of the effective action by exploiting the (background)
gauge invariance, which in general yields Ward identities, i.e. 
linear relations among the vertex functions. This is a noticeable
technical advantage in comparison with the Slavnov-Taylor identities.

Thus the construction of the background effective action turns out
to be simpler than the quantization of the ordinary underlying
gauge theory, since the structure of counterterms can be greatly
simplified by using the symmetry requirement of background gauge invariance.
For this reason the background field method has been advantageously applied
to gravity and supergravity computations \cite{Gates:1983nr} and  to
the calculation
of the $\beta$ function in Yang-Mills theory \cite{abbott1,betaQCD}.
More recently, it has been applied to the quantization of
the Standard Model \cite{denner,grassi}.

Recent applications of the background field method
in the context of the renormalization group flow \cite{Becchi:2000bq}
suggest to construct first a regularized background-gauge invariant
effective action and then to recover the
Slavnov-Taylor identities (broken by the regularization procedure)
by fine-tuning the free parameters of the model.

The quantization of gauge theories in perturbative quantum
field theory requires
the introduction of a gauge-fixing term.
The gauge-invariance of the effective action is thus
spoiled even at the classical level. Gauge invariance is only
recovered at the very end of the calculations, when physical amplitudes
are considered.

The classical gauge-fixed action is however 
symmetric under the BRST transformations
and for non-anomalous theories, the corresponding quantum action satisfies the
Slavnov-Taylor (ST) identities.

The introduction of the background field method quantization
allows to define a modified effective action,
the background effective action, depending
on classical background gauge-fields and matter-fields, which,
in the non-anomalous cases, can be chosen to be symmetric
under suitably defined linear background gauge transformations.

Already in early papers on the background field method, as for
instance in Ref.~\cite{abbott2}
and more recently in Refs. \cite{grassi1,grassi2,becchi} it
has been pointed out that  a further gauge fixing
term is needed beside the usual background gauge.
The seminal discussion of \cite{abbott2} has been improved
in \cite{becchi}, where the problem of the extra gauge-fixing term,
needed to construct connected amplitudes,
has been clarified.
In fact the construction of the connected amplitudes by using the
effective action, obtained with the background field method, fails
in the case of the two-point function, since in this case the vertex
function has no inverse. In the present approach this new gauge fixing
term is introduced right at the beginning in the action and it
is chosen in such a way
that BRST invariance is preserved, although the gauge invariance
of the background effective action is lost. This difficulty is overcome
by defining a new background effective action by subtracting a harmless
functional linear in the quantized fields. Linearity is important 
in order to avoid the introduction of a {\sl ad hoc}
composite operator.

The interplay of the background field symmetry with Slavnov-Taylor
identities
has first been considered in Ref.~\cite{abbott2}, where the
equivalence of the background gauge field quantization
and the ordinary perturbative quantization of gauge theories was proven
by using a Ward identity derived from BRST invariance of the
theory.
It turns out \cite{grassi1} to be particularly useful to allow the
background fields to vary under the BRST transformation $s$,
so that the background fields enter
into the BRST transformations as doublets, together with a suitably
defined set of classical ghosts $\Omega$.
For instance the transformation of
the background gauge-fields $V^\mu_a$ are
\begin{eqnarray}
s V^\mu_a = \Omega^\mu_a \, , ~~~~ s \Omega^\mu_a = 0 \, .
\label{introq1}
\end{eqnarray}
The introduction of the doublet $(V^\mu_a,\Omega^\mu_a)$
allows an elegant derivation of the Slavnov-Taylor identity
for the connected amplitudes used in the
background equivalence theorem. It is however necessary
to show that the extension of the BRST transformations given by eq.
(\ref{introq1}) does not alter the anomaly content of the theory.
In fact the removal of the doublet has to be explicitly checked in
the background field method,
as it is done in Ref.~\cite{grassi1} for $SU(n)$
and with emphasis on the Standard Model group in Ref.~\cite{grassi,grassi1bis}.
In the proof
presented in this paper we do not assume power-counting: only
locality of the breaking term of the Slavnov-Taylor identity is
necessary.

The doublet $(V^\mu_a,\Omega^\mu_a)$ is not
essential for the construction of the theory. In fact the physical
content of the theory is insensitive to the BRST transformation
properties of  $V^\mu_a$. As an example, consider the alternative
approach where $s V^\mu_a = 0$ and compare with the case described
by eq.(\ref{introq1}) for the local term $F_a^{\mu\nu}(V)^2$.
In the first case this term is a ST-invariant and therefore can be
removed from the action by a suitable normalization condition
\cite{ferrari1,ferrari2}, while in the second case it is removed
by the requirement of ST-invariance.

In the previous approaches \cite{abbott2,becchi}
emphasis was put on the independence
from the background gauge of the physical elements of the S-matrix.
For this reason the discussion has been limited \cite{becchi} to
an infrared-safe theory. In fact the result can be extended to quite a
general
situation where one considers any expectation value of quasi-local
observable operator (therefore BRST invariant objects), thus including
the physical S-matrix elements of a massive theory. Thus we consider
physical observables given by the equivalent classes of ST invariant
quasi-local operators, where the equivalence relation is given by
\begin{eqnarray}
{\cal O}\sim {\cal O'} + s X
\label{introq2}
\end{eqnarray}
for some $X$.

In the present paper we consider also the Legendre transform
of the background effective action. Thus we get the functional 
$W_{\rm bg}$,
which generates the connected amplitudes. By this procedure 
one needs to derive the free propagator of the gauge field
from the background effective action,
 which exists only if an extra gauge fixing term,
beside the background one, as it has been stressed previously
\cite{becchi}. By using the relevant ST identities
we show that the functional $W_{\rm bg}$ gives the same
physical amplitudes as the original one we started with.
Moreover we show that $W_{\rm bg}$ in general cannot
be derived from a classical action by the Gell-Mann-Low
formula. In fact the Feynman rules for the vertices inside
a 1PI amplitude and for those containing a connecting
line are different in general (not for abelian gauge theories). 

The paper is organized as follows.
%
%
Section~\ref{sec:basic} is devoted to
a short summary of the background field method.
Here the BRST is extended to the background gauge-fields $V$
as in previous discussions (see 
Refs.~\cite{grassi1,grassi2,becchi}).
This provides a technical advantage in
the derivation of the necessary Slavnov-Taylor identities.
Section~\ref{sec:bfm} shows the independence of the physical amplitudes
from the background field.
Section~\ref{sec:how} extends the theorem of equivalence
for the background field method to the general case
of vacuum expectation value of physical observables,
and deals with a thorough discussion of the
background field method construction of  physical amplitudes.
Section~\ref{sec:bg} shows the properties of the Legendre transform of
the background effective action and the physical equivalence of the
background functional for the connected amplitudes. 
Section~\ref{sec:doublets} contains an improved proof
that the extended BRST transformation on the background
gauge field $V$ (eq.(\ref{introq1})) does not change the anomaly properties
of the Slavnov-Taylor identity. The conclusions in 
Section~\ref{sec:conclusions} provide
a summary of the background field method and the improvements
of the method obtained in the present paper.

\section{The generating functional for the Feynman amplitudes}
\label{sec:basic}
We consider a generic gauge theory; however matter fields will
not be displayed in the notations. The fields
are $A_{a\mu}$ (gauge fields), $B_a$ (Lagrange multipliers in the
gauge fixing), $V_{a\mu}$ (the background fields), $\Omega_{a\mu}$
(the BRST partner of $V_{a\mu}$), ${\bar \theta}_a, \theta_a$
(the Faddeev-Popov fields).
The action is
\begin{eqnarray}
\back\Gamma^{(0)}[A,V] &=&\int d^4 x\left\{ -\frac{1}{4}F^2_{\mu\nu}(A)
+\alpha ~ s\,\left[{\bar \theta}_a\left(\frac{B_a}{2}-f_a\right)\right]
-\alpha' s\, \left[{\bar \theta}_a\partial^\mu V_{a\mu}\right]\right\}
\label{ab1}
\end{eqnarray}
where $s$ is the BRST operator and
\begin{eqnarray}
f_a = \partial_\mu(A-V)^{\mu}_a +f_{abc}V^\mu_b(A-V)_{c\mu}
\equiv D^\mu(V)_{ac}(A-V)_{c\mu}
\label{ab2}
\end{eqnarray}
is the background gauge fixing  in the notation
\begin{eqnarray}
D^\mu(V)_{ac}\equiv \delta_{ac}\partial^\mu+
f_{abc}V^\mu_b.
\label{ab3}
\end{eqnarray}
The BRST transformations are
\begin{eqnarray}
\begin{array}{lcl}
s\, A_{a\mu} = D_\mu(A)_{ab}\theta_b
&
{~~~~~~~}
&
s\, \theta_a = - \frac{1}{2} f_{abc}\theta_b\theta_c
\\
s\, {\bar \theta}_a = B_a
&
{~~~~~~~}
&
s\, B_a = 0
\\
s\, V_{a\mu} = \Omega_{a\mu}
&
{~~~~~~~}
&
s\, \Omega_{a\mu} = 0.
\end{array}
\label{ab4}
\end{eqnarray}
The $\alpha'$ term is introduced in order to deal with
the degenerate case $A=V$ (i.e. $f=0$). This will be discussed
later on in Section~\ref{sec:bg}.
\par
By using the transformations (\ref{ab4}) the action becomes
\begin{eqnarray}
\Gamma^{(0)}[A,V] &=&
\int d^4 x\Big\{ -\frac{1}{4}F^2_{\mu\nu}(A)
\nonumber\\&&
+\alpha ~\left(\frac{B^2}{2}-BD(V)(A-V)+{\bar \theta}\left( D(V)D(A)\theta
-D(A)\Omega \right)\right)
\nonumber\\&&
-\alpha'\left( B\partial^\mu V_\mu - \bar\theta\partial^\mu \Omega_\mu
\right)
+ \bar\theta^* B
+ A^*D(A)\theta
- \frac{1}{2}\theta^*_a f_{abc}\theta_b\theta_c
\Big\}.
\label{ab5}
\end{eqnarray}
The anti-fields $A^*,\bar\theta^*,\theta^* $ are the external
fields coupled to the BRST-transforms
($D_\mu(A)_{ab}\theta_b$, $B_a$, $-\frac{1}{2} f_{abc}\theta_b\theta_c$)
 of the fundamental fields.
\par
Now we consider the generating functional of the Green functions
associated to the above action, with external currents $J,K,\eta,\bar\eta$
coupled to the quantized fields $A,B,\bar \theta,\theta$:
\begin{eqnarray}
Z[J,V,\dots] &=&\exp\left(iW\right)\equiv
\int {\cal D}A{\cal D}B{\cal D}{\bar \theta}
{\cal D}\theta
\exp i\int d^4 x\left\{ -\frac{1}{4}F^2_{\mu\nu}(A)\right.
\nonumber \\ &&\quad
+\alpha ~ \left(\frac{B^2}{2}-Bf+{\bar \theta}\left( D(V)D(A)\theta
-D(A)\Omega\right)\right)
\nonumber \\ &&\quad
-\alpha'\left( B\partial^\mu V_\mu - \bar\theta\partial^\mu
\Omega_\mu\right)
\nonumber \\ &&\quad
\left.
+ JA+KB - \eta{\bar \theta}
-{\bar\eta}\theta
+A^*D(A)\theta - \frac{1}{2}\theta^*_a f_{abc}\theta_b\theta_c
\right\}\, .
\label{ab6}
\end{eqnarray}
\par
The invariance under the BRST transformations corresponds
to require the validity of the Slavnov-Taylor identity
\begin{eqnarray}
{\cal S} W[J,V,\dots]&=&\int d^4 x
\left ( - J \frac{\delta}{\delta A^*}
-  \eta \frac{\delta}{\delta \bar\theta^*}
- {\bar\eta} \frac{\delta}{\delta \theta^*} 
+ \Omega \frac{\delta}{\delta V} \right) W[J,V,\dots]
\nonumber\\
&=&0
\label{ab7a}
\end{eqnarray}
for the generating functional
of the (connected) Green functions.
The anti-field $\bar\theta^*_a$
is the external field coupled to the BRST-transforms $B$
of the fundamental field $\bar\theta$,
then the current $K_a$ can be identified with $\bar\theta^*_a$.
The Feynman rules of the perturbative expansion can be read from 
eq.(\ref{ab6}).
\subsection{Background gauge symmetry}
For $\alpha' =0$ the action in eq.(\ref{ab5}) is invariant under
the {\sl background gauge transformations}.
\begin{eqnarray}
\begin{array}{lcl}
\delta\, A = D(A)\omega
&
{~~~~~~~}
&
\delta\, V = D(V)\omega
\\
\delta\, \phi_a = -f_{abc}\omega_b\phi_c
&
{~~~~~~~}
&
\mbox{with }
\phi_a\in\{\theta_a,\bar\theta_a,B_a,\Omega_a, A^*_{a\mu},\theta^*_a\}
\end{array}
\label{ab7}
\end{eqnarray}
where $\omega_a$ is a group parameter. Consequently the functional
\begin{eqnarray}
{\tilde Z}[J,V,\dots]=\exp\left(i\tilde W\right)\equiv
Z[J,V,\dots]~\exp\left\{-i\int d^4x V J\right\}
\label{ab8}
\end{eqnarray}
is, for $\alpha' =0$, invariant under the transformation
\begin{eqnarray}
\begin{array}{lcl}
\delta\, V = D(V)\omega
&
{~~~~~~~}
&
\delta\, \Omega_{a\mu} = -f_{abc}\omega_b \Omega_{c\mu}
\\
\delta\, \zeta_a = -f_{abc}\omega_b \zeta_{c}
&
{~~~~~~~}
&
\mbox{with }
\zeta_a\in\{J_{a\mu}, K_a, {\bar\eta}_a, \eta_a, A^*_{a\mu}, \theta^*_a\}.
\end{array}
\label{ab9}
\end{eqnarray}
The corresponding Ward identity
\begin{eqnarray}
\back{\cal G}_a(x){\tilde W }&\equiv&
\left\{-D_\mu(V)_{ab}\frac{\delta}{\delta V_{b\mu}(x)}
- f_{abc} \left( \Omega_{b\mu}(x) \frac{\delta}{\delta \Omega_{c\mu}(x)}
+ J_{b\mu}(x)\frac{\delta}{\delta J_{c\mu}(x)}
\right.\right.
\nonumber \\ &&
\left.\left.
+ K_{b}(x)\frac{\delta}{\delta K_{c}(x)}
+\eta_{b}(x)\frac{\delta}{\delta \eta_{c}(x)}
+{\bar\eta}_{b}(x)\frac{\delta}{\delta {\bar\eta}_{c}(x)}
+A^*_{b\mu}(x)\frac{\delta}{\delta A^*_{c\mu}(x)}
\right.\right.
\nonumber \\ &&
\left.\left.
+\theta^*_{b}(x)\frac{\delta}{\delta \theta^*_{c}(x)}
\right)\right\}{\tilde W }=0
\label{ab9a}
\end{eqnarray}
can be used in the renormalization procedure 
together with the ST (derived from eq.(\ref{ab7a}))
\begin{eqnarray}
\back{\cal S}\tilde W[J,V,\dots]&=&\int d^4 x\left\{
\left ( - J\frac{\delta}{\delta A^*}
-  \eta \frac{\delta}{\delta \bar\theta^*}
- {\bar\eta} \frac{\delta}{\delta \theta^*} 
+ \Omega \frac{\delta}{\delta V} \right)\tilde W[J,V,\dots]
-
 J\Omega\right\}
\nonumber\\&=&0\,,
\label{ab7aa}
\end{eqnarray}
since the corresponding operators commute
\begin{eqnarray}
\left [ {\cal S},{\cal G}_a\right ] =0\,.
\label{ab9ap}
\end{eqnarray}
\par
If $\alpha' \not=0$ the breaking of the {\sl background
gauge symmetry} is harmless since it is due to
a gauge fixing term linear in the quantized fields. The associated
Ward identity for the functional of the connected amplitudes
$\tilde W$ is then modified by a term proportional to $\alpha'$
\begin{eqnarray}
&&\back{\cal G}^{(\alpha')}_a(x){\tilde W }\equiv
\left\{-D_\mu(V)_{ab}\frac{\delta}{\delta V_{b\mu}(x)}
- f_{abc} \left( 
\Omega_{b\mu}(x) \frac{\delta}{\delta \Omega_{c\mu}(x)} 
+J_{b\mu}(x)\frac{\delta}{\delta J_{c\mu}(x)}
\right.\right.
\nonumber \\&&
\left.\left.
+ K_{b}(x)\frac{\delta}{\delta K_{c}(x)}
+\eta_{b}(x)\frac{\delta}{\delta \eta_{c}(x)}
+{\bar\eta}_{b}(x)\frac{\delta}{\delta {\bar\eta}_{c}(x)}
+A^*_{b\mu}(x)\frac{\delta}{\delta A^*_{c\mu}(x)}
\right.\right.
\nonumber \\&&
\left.\left.
+\theta^*_{b}(x)\frac{\delta}{\delta \theta^*_{c}(x)}
\right)
+ \alpha' \partial^\mu \left(
D_\mu(V)_{ab}\frac{\delta}{\delta K_{b}(x)}
-f_{abc}\Omega_{b\mu}\frac{\delta}{\delta \eta_{c}(x)}
\right)\right\}{\tilde W }
=0\,.
\label{ab9b}
\end{eqnarray}
By explicit computation one can verify that also
\begin{eqnarray}
\left [ {\cal S},{\cal G}_a^{(\alpha')}\right ] =0\,.
\label{ab9app}
\end{eqnarray}
Then the renormalization program can be performed by using
both conditions expressed by the ST identity in eq.(\ref{ab7aa})
and the Ward identity given in eq.(\ref{ab9a}) or eq.(\ref{ab9b}).
%
\subsection{Gauge invariant effective action}

It is worthwhile to illustrate from a different point of view the
properties of the functionals $\tilde Z$.
By a simple change of variable in the functional integral one gets
\begin{eqnarray}
\back{\tilde Z}[J,V,\dots] &=&\exp(iW) =
\int {\cal D}{\tilde A}{\cal D}B{\cal D}{\bar \theta}
{\cal D}\theta
\exp i\int d^4 x\Big\{ -\frac{1}{4}F^2_{\mu\nu}({\tilde A}+V)
\nonumber \\ &&
+\alpha ~ \Big(\frac{B^2}{2}-BD^\mu(V){\tilde A}_\mu
+{\bar \theta}\big( D(V)D({\tilde A}+V)\theta
-D({\tilde A}+V)\Omega \big)\Big)
\nonumber \\ &&
-\alpha'\left( B\partial^\mu V_\mu - \bar\theta\partial^\mu
\Omega_\mu\right)
\nonumber \\ &&
+ J{\tilde A}+KB - \eta{\bar \theta}
-{\bar\eta}\theta
+A^*D(\tilde A+V)\theta - \frac{1}{2}\theta^*_a f_{abc}\theta_b\theta_c
\Big\}.
\label{ab6.p}
\end{eqnarray}
This expression gives the Feynman rules for any Green function
involving any number of derivatives with respect to $V$. In fact the
functional dependence of ${\tilde Z}$ from $V$ should be
understood as given by the ensemble of all possible
derivatives of ${\tilde Z}$ with respect to  $V$.
Now we can introduce the generating functional for the 1PI 
Green functions 
associated to $\tilde Z$,
i.e. we set
\begin{eqnarray}
{\tilde A} = \frac{\delta}{\delta J}{\tilde W}
\label{ab11}
\end{eqnarray}
and perform the Legendre transform
\begin{eqnarray}
{\tilde \Gamma}[{\tilde A},V]  = {\tilde W} -
\int d^4 x J {\tilde A} - \dots
\label{ab11.1}
\end{eqnarray}
In eq.(\ref{ab11.1}) $\tilde\G$ is the full Legendre transform of
$\tilde W$. Dots indicate the remaining conjugate variables besides
($J,\tilde A$).
By using eq.(\ref{ab8}) one gets
\begin{eqnarray}
{\tilde A}
=\frac{\delta}{\delta J}W -V
=A-V\,,
\label{ab11p}
\end{eqnarray}
and therefore
\begin{eqnarray}
{\tilde \Gamma}[{\tilde A},V]  &=& {\tilde W} -
\int d^4 x J{\tilde A} - \dots
\nonumber\\
 &=& W - \int d^4 x J ({\tilde A}+V) - \dots
\nonumber\\
 &=& \Gamma[A,V]_{\big|{A={\tilde A}+V}}\,.
\label{ab11.1p}
\end{eqnarray}
The lowest order ${\tilde \Gamma} $ is given by (see eq.(\ref{ab5}))
\begin{eqnarray}
{\tilde\Gamma}^{(0)}[{\tilde A},V] &=&
\int d^4 x\Big\{ -\frac{1}{4}F^2_{\mu\nu}({\tilde A}+V)
\nonumber\\&&
+\alpha ~ \left(\frac{B^2}{2}-BD(V){\tilde A}+{\bar \theta}
(D(V)D({\tilde A}+V)\theta - D(\tilde A + V)\Omega)\right)
\nonumber\\&&
-\alpha'\left( B\partial^\mu V_\mu - \bar\theta\partial^\mu
\Omega_\mu\right)
+A^*D(\tilde A+V)\theta - \frac{1}{2}\theta^*_a f_{abc}\theta_b\theta_c
\Big\}.
\label{ab12}
\end{eqnarray}
The generating functional of the 1PI $\tilde \Gamma$ associated
to ${\tilde W }$ satisfies the corresponding Ward identity
in eq.(\ref{ab9b})
\begin{eqnarray}
&&\back{\cal G}^{(\alpha')}_a(x)\tilde\Gamma[\tilde A,V]=
\Big\{
- D_\mu(V)_{ab}\frac{\delta}{\delta V_{b\mu}(x)}
- f_{abc} \Big[
    \Omega_{b\mu}(x) \frac{\delta}{\delta \Omega_{c\mu}(x)}
\nonumber\\&&
  + \tilde A _{b\mu}(x)\frac{\delta}{\delta \tilde A_{c\mu}(x)}
  + B_{b}(x)\frac{\delta}{\delta B_{c}(x)}
  + {\bar\theta}_{b}(x)\frac{\delta}{\delta {\bar\theta}_{c}(x)}
  + \theta_{b}(x)\frac{\delta}{\delta \theta_{c}(x)}
\nonumber\\&&
  + A^*_{b\mu}(x)\frac{\delta}{\delta A^*_{c\mu}(x)}
  + \theta^*_{b}(x)\frac{\delta}{\delta \theta^*_{c}(x)}
\Big]
\Big\}\tilde\Gamma[\tilde A,V]
\nonumber \\ &&
+\alpha'\partial^\mu \left(D_\mu(V)_{ab}B_b
+ f_{abc}\bar\theta_b\Omega_{c\mu}
\right)=0
\label{ab9cppp}
\end{eqnarray}
and the ST identity
\begin{eqnarray}
\int &d^4x& \,
\left \{
\Omega \frac{\delta \tilde \Gamma[{\tilde A},V]}{\delta V}+
B \frac{\delta\tilde \Gamma[{\tilde A},V]}{\delta\bar \theta}+
\frac{\delta\tilde \Gamma[{\tilde A},V]}{\delta\theta^*}
\frac{\delta\tilde \Gamma[{\tilde A},V]}{\delta\theta}
\right.\nonumber\\&&\left.
 - \left(-\frac{\delta\tilde \Gamma[{\tilde A},V]}{\delta A^*}+\Omega
\right)
\frac{\delta\tilde \Gamma[\tilde A,V]}{\delta\tilde A}\right \}=0\,.
\label{ab21q}
\end{eqnarray}
The $\alpha'$ part in eq.(\ref{ab9cppp}) can be easily removed
by introducing the functional
\begin{eqnarray}
\Gamma_{\rm bg}[\tilde A,V] \equiv \tilde \Gamma[\tilde A,V] +
\alpha'\int d^4x\Big\{
 B\partial^\mu V_\mu - \bar\theta\partial^\mu \Omega_\mu\Big\}.
\label{ab10a}
\end{eqnarray}
It is straightforward to show that $\Gamma_{\rm bg}[\tilde A,V]$
satisfies the Ward identities
\begin{eqnarray}
&&\back{\cal G}_a(x)\Gamma_{\rm bg}[\tilde A,V]=
\Big\{
- D_\mu(V)_{ab}\frac{\delta}{\delta V_{b\mu}(x)}
- f_{abc} \Big[
    \Omega_{b\mu}(x) \frac{\delta}{\delta \Omega_{c\mu}(x)} 
\nonumber\\&&
  + \tilde A _{b\mu}(x)\frac{\delta}{\delta \tilde A_{c\mu}(x)}
  + B_{b}(x)\frac{\delta}{\delta B_{c}(x)}
  + {\bar\theta}_{b}(x)\frac{\delta}{\delta {\bar\theta}_{c}(x)}
  + \theta_{b}(x)\frac{\delta}{\delta \theta_{c}(x)}
\nonumber \\ &&
  + A^*_{b\mu}(x)\frac{\delta}{\delta A^*_{c\mu}(x)}
  + \theta^*_{b}(x)\frac{\delta}{\delta \theta^*_{c}(x)}
\Big]
\Big\}\Gamma_{\rm bg}[\tilde A,V]
%
=0\ .
\label{ab9c}
\end{eqnarray}
These Ward identities follow from the invariance of
$\Gamma_{\rm bg}[\tilde A,V]$ under the transformation
\begin{eqnarray}
\begin{array}{lcl}
\delta\, {\tilde A}_{a\mu} =  -f_{abc}\omega_b {\tilde A}_{c\mu}
&
{~~~~~~~}
&
\delta\, V = D(V)\omega
\\
\delta\, \phi_a = -f_{abc}\omega_b\phi_c
&
{~~~~~~~}
&
\mbox{with }
\phi_a\in\{\theta_a,\bar\theta_a,B_a,\Omega_a, A^*_{a\mu},\theta^*_a\},
\end{array}
\label{ab13}
\end{eqnarray}
while the generating functional of the 1PI $\tilde \Gamma$
is not invariant under the {\sl background
gauge transformations } due to the breaking term proportional
to $\alpha'$.
%
%
%
\subsection{Further algebra}
There is an interesting limit: take $\tilde A \to 0$. Then from 
eq.(\ref{ab11.1}) one gets
\begin{eqnarray}
{\tilde \Gamma}[0,V,\dots]  = {\tilde W}[J,V,\dots]-\int d^4x \left(
KB - \eta{\bar \theta}-{\bar\eta}\theta
\right)
\label{ab14}
\end{eqnarray}
where $J$ is given by eq.(\ref{ab11})
\begin{eqnarray}
{\tilde A} = \frac{\delta}{\delta J}{\tilde W}[J,V,\dots] = 0.
\label{ab15}
\end{eqnarray}
Moreover one gets
\begin{eqnarray}
{\tilde \Gamma}[0,V]  = \Gamma[V,V].
\label{ab16}
\end{eqnarray}
This very interesting equation looks very simple and innocuous.
In fact the LHS is given by the 1PI amplitudes constructed with
the Feynman rules given by the action in eq.(\ref{ab12}), while
the RHS is given by the 1PI amplitudes provided by the action
used to define the generating functional 
$Z$ in eq.(\ref{ab6}).
%
%
%
\section{Independence of the physical amplitudes from the background field}
\label{sec:bfm}
Physical amplitudes should be independent from the background
field $V_{a\mu}$, from the source $K_a$ of the field $B_a$ and from the
gauge parameters $\alpha$ and $\alpha'$.
This can be proved by using the Slavnov-Taylor identities~(\ref{ab7a})
introduced in Section~\ref{sec:basic}.
\par
The physical amplitudes can be obtained by introducing a set
of external sources $\beta_i(x),\quad i=1,\dots$ coupled to
local or quasi-local BRST invariant quantities. Both $A_{a\mu}$ and $B_a$
cannot be physical fields, therefore $J_{a\mu}$ and $K_a$ cannot belong to the
set of $\beta$ sources. Let us consider the dependence
of the physical amplitudes on $V$, by taking the derivative
of eq.(\ref{ab7a}) with respect to $\Omega$ and then putting
$\eta=\bar\eta=\Omega=0$
\begin{eqnarray}
\Big\{- \int d^4 y J(y)\frac{\delta^2}{\delta\Omega(x)\delta A^*(y)}
+\frac{\delta}{\delta V(x)}\Big\}W[J,V,\dots]_{\big|
\eta=\bar\eta=\Omega=0}=0\,.
\label{ad4}
\end{eqnarray}
Notice that the insertion of the operator coupled to $\Omega_{b\mu}$
\begin{eqnarray}
- \alpha D_\mu(A)_{bc}{\bar\theta}_c + \alpha'\partial_\mu{\bar\theta}_b
\label{ad5}
\end{eqnarray}
is just (for $\alpha'=0$) the composite operator appearing in 
eq.(2.10) of Ref.~\cite{abbott2}.
Eq.(\ref{ad4}) is the starting point to prove the independence
of physical amplitudes from $V$. Any number of derivatives can
be taken with respect to $\beta_i(x)$ and moreover one has to
put $J=0$ being conjugated to an unphysical field. The result
is independent from the value of $V$. 
\par
A similar argument can be used in order to prove
that the derivative with respect to the source $K$ of any
physical amplitude is zero.
In fact by taking the functional derivative with respect to $\eta$
of eq.(\ref{ab7a}) one gets
\begin{eqnarray}
\Big\{-\int d^4 yJ(y)\frac{\delta^2}{\delta\eta(x)\delta A^*(y)}
-\frac{\delta}{\delta {\bar\theta^*(x)}}\Big\}
W[J,V,\dots]_{\big|\eta=\bar\eta=\Omega=0}=0
\label{ad6}
\end{eqnarray}
and the derivative with respect to $\bar\theta^*$ takes down
a $B$ field. Thus there is no $K$-dependence for the physical
amplitudes even in presence of the external fields $V$.
\par
The independence of the physical amplitudes from $\alpha$
and $\alpha'$ can be established according to the standard
arguments~\cite{piguet}. The proof follows the same pattern
as the one outlined above.
\section{The background equivalence theorem}
\label{sec:how}
This section reports a well known result and it is included
only to make the discussion self contained.

The results of the previous Sections allow us
to formulate the theorem of equivalence for the
background field method in a rather simple way:
\par\noindent
{\bf Theorem of equivalence}: The construction of
the connected amplitudes functional can be
equivalently performed (i.e. giving the same physical amplitudes)
either by using the 1PI vertex  amplitudes generated by the 
 functional $\Gamma[A,V]$
or by using the 1PI vertex 
 amplitudes 
generated by $\tilde\Gamma[0,V]|_{V=A}$. 
\par
We remark that the functional $\tilde\Gamma[0,V]$ gives the same
 1-PI \   functions as the gauge invariant functional
$\Gamma[0,V]_{\rm bg}$ except for the two-point functions.
\par
The proof is as follows \cite{abbott2}.
The starting theory is described by the classical action
in eq.(\ref{ab1}). Correspondingly, the functionals one has to compute
are  $\Gamma[A,V]$ and $W[J,V]$. If  $\Gamma[A,V]$ is known then the
connected amplitudes, generated by $W[J,V]$, can be obtained
by gluing together the 1PI vertex amplitudes with the connected
two-point functions provided by the inverse of
the relevant part of $-\Gamma[A,0]|_{\beta=0}$.

We now restrict ourselves to physical quantities.
Eq.(\ref{ad4})
says that
\begin{eqnarray}
\left . \frac{\delta^{(n+1)} W}{\delta V\delta \beta_1\dots \delta\beta_n}
\right |_{J=\eta=\bar\eta=0}=
\left . \frac{\delta^{(n+1)} \Gamma}{\delta V\delta \beta_1\dots \delta\beta_n}
\right |_{\frac{\delta \G}{\delta A} = 
\frac{\delta \G}{\delta \bar \theta} =
\frac{\delta \G}{\delta \theta}=0 }=0\,,
\label{ab14bp}
\end{eqnarray}
then, in constructing the connected amplitude for physical
processes, one can add to any 1PI vertex
$\Gamma_{A_{{a_1}{\mu_1}}\dots A_{{a_n}{\mu_n}}}[A,V,\beta]|_{A=0}$
all those, where some derivatives with respect to $A_{b\nu}$ are replaced
by the corresponding derivatives with respect to $V_{b\nu}$.
This procedure amounts to replace $\Gamma[A,V]$ with
$\Gamma[A,A]$ i.e. with $\tilde\Gamma[0,V]|_{V=A}$ according to
eq.(\ref{ab16}). In general the resulting connected amplitudes
will be different from those generated by $W$; however the
physical amplitudes will coincide. This concludes the proof.
\par
We would like to stress that the connected two-point functions
used in the above construction are  the ones generated by
$\Gamma[A,V]$ expanded in powers of $V$  and not the ones generated by
$\tilde\Gamma[0,V]|_{V=A}=\Gamma[A,A]$. Thus the background equivalence
theorem is valid also in the case $\alpha'=0$.

%
%
\section{The background gauge functional for connected amplitudes}
\label{sec:bg}

In the construction outlined in sect.\ref{sec:how} 
the 1-PI vertex amplitudes depending on $V_{a\mu}$ are
joined by the connected two-point functions generated by
$\Gamma[A,0]$.
This is the choice
adopted in \cite{abbott1,abbott2,becchi}.

However, this 
is not strictly necessary \cite{becchi}. 
I.e. the physical amplitudes can be obtained 
from the connected Green functions  where the 1-PI vertex amplitudes   
$\tilde\Gamma[0,V]|_{V=A}$ are glued together
by using connected
two-point functions different from those used in 
sect.\ref{sec:how}. One interesting possibility is provided
by introducing the Legendre transform of the vertex functional
$\tilde\Gamma[0,V]|_{V=A}$.
In this case 
one needs the extra gauge fixing parameter $\alpha'\not =0$
in order that the bilinear part of  
$\tilde\Gamma[0,V]|_{\beta=0}$
possesses the inverse.

\subsection{Legendre transform of the background effective action}

Then one can introduce a new functional
$W_{\rm bg}$ defined by
%
\begin{equation}
\back\left\{
\begin{array}{l}
{\displaystyle
W_{\rm bg}[J,\beta,\dots] =
\tilde\Gamma[0,V,\beta,\dots]+\int d^4 x \,J\, V 
+\int d^4x \left(
KB - \eta{\bar \theta}-{\bar\eta}\theta
\right)
} \\
{\displaystyle
J= - \frac{\delta \tilde\Gamma[0,V]}{\delta V}
} \\
{\displaystyle
K = -\frac{\delta \tilde \Gamma[0,V]}{\delta B}\, ,
\eta =  -\frac{\delta \tilde \Gamma[0,V]}{\delta \bar\theta} \, ,
\bar \eta =  -\frac{\delta \tilde \Gamma[0,V]}{\delta \theta} \, .
}\end{array}\right .
\label{ab9cpp}
\end{equation}
The functional derivatives of $W_{\rm bg}$ with respect to the sources
$\beta$ reproduce in the on-shell limit $J=\beta=0$ the on-shell
physical connected amplitudes of the original theory.
We will report a proof of this fact in sect.\ref{sec:pheq}.

The construction of $W_{\rm bg}$ is possible only for
$\alpha'\not =0$, since, otherwise, the bilinear part of 
$\tilde\Gamma[0,V]$ does not possess the inverse.
It is important to notice that in general $W_{\rm bg}$
will not correspond to a field theory. In fact the Feynman
rules for vertices involving only internal legs in the 1PI
graphs ($\tilde A$ vertices in eq.(\ref{ab12}))
do not coincide with those, where some $V$ are involved.
See the rules given by the free action
in eq.(\ref{ab12}).

As a consequence of these facts one
cannot derive identities for $W_{\rm bg}$ and for
$\tilde\Gamma[0,V]$ from invariance properties of the
{\em classical} action (action principle). We have to revert to the properties
of $\tilde\Gamma[A,V]$ and $\tilde W$ and use eqs.(\ref{ab7aa})
and (\ref{ab9cppp}).
From eq.(\ref{ab9cppp})
\begin{eqnarray}
&&\back{\cal G}^{(\alpha')}_a(x)\tilde\Gamma[0,V]=
\Big\{
- D_\mu(V)_{ab}\frac{\delta}{\delta V_{b\mu}(x)}
- f_{abc} \Big[
   \Omega_{b\mu}(x) \frac{\delta}{\delta \Omega_{c\mu}(x)} 
\nonumber \\ &&
  + B_{b}(x)\frac{\delta}{\delta B_{c}(x)}
  + {\bar\theta}_{b}(x)\frac{\delta}{\delta {\bar\theta}_{c}(x)}
  + \theta_{b}(x)\frac{\delta}{\delta \theta_{c}(x)}
  + A^*_{b\mu}(x)\frac{\delta}{\delta A^*_{c\mu}(x)}
\nonumber \\ &&
  + \theta^*_{b}(x)\frac{\delta}{\delta \theta^*_{c}(x)}
\Big]
\Big\}\tilde\Gamma[0,V]
+\alpha'\left(D_\mu(V)_{ab}B_b
+ f_{abc}\bar\theta_b\Omega_{c\mu}
\right)=0\,,
\label{ab99c}
\end{eqnarray}
which corresponds to setting $\tilde A_{a\mu}=0$ in the transformations
in eq.(\ref{ab13}).
\par
In eq.(\ref{ab7aa}) one can put $J=\bar J[\tilde A,V]$, the solution
of eq.(\ref{ab11p})
\begin{eqnarray}
\back{\cal S}\tilde W[\bar J,V,\dots]&=&\int d^4 x
\left ( -\bar J \frac{\delta}{\delta A^*}
- \eta \frac{\delta}{\bar \eta^*}
- {\bar\eta} \frac{\delta}{\delta \theta^*}
+ \Omega \frac{\delta}{\delta V} \right)\tilde W[\bar J,V,\dots] 
\nonumber\\&&
-\int d^4 x \bar J[\tilde A,V]\Omega=0
\label{ab7aap}
\end{eqnarray}
which can be written in terms of $\tilde\Gamma[0,V]$ setting $\tilde A=0$:
\begin{eqnarray}
\back\int d^4 x\left\{
\left (-\bar J[0,V] \frac{\delta}{\delta A^*}
-  {\bar\eta} \frac{\delta}{\delta \theta^*}
+ \Omega \frac{\delta}{\delta V} \right)\tilde \Gamma[0,V]
- \eta B - \bar J[0,V]\Omega\right\}=0\,.
\label{ab7aapp}
\end{eqnarray}
The above equation shows that the ST identity does not close,
in fact from eq.(\ref{ab11.1p}) we have
\begin{eqnarray}
\bar J[0,V]=
- \frac{\delta \tilde\Gamma[\tilde A,V]}{\delta \tilde A}|_{\tilde A=0}\,,
\label{ab7q}
\end{eqnarray}
and finally we get
\begin{eqnarray}
&&\back\int d^4x \,
\left \{
\Omega \frac{\delta \tilde \Gamma[0,V]}{\delta V}+
B \frac{\delta\tilde \Gamma[0,V]}{\delta\bar \theta}+
\frac{\delta\tilde \Gamma[0,V]}{\delta\theta^*}
\frac{\delta\tilde \Gamma[0,V]}{\delta\theta}
\right \} =
\nonumber\\&&
~~~~~~~~~~~~~~~~~~~~~~~~~~\int d^4x \,
 \left(-\frac{\delta\tilde \Gamma[0,V]}{\delta A^*}+\Omega
\right)
\left ( \frac{\delta\tilde \Gamma[\tilde A,V]}{\delta\tilde A}
\right )_{\tilde A=0}.
\label{ab21}
\end{eqnarray}
Therefore in general the ST identity cannot be satisfied
by $\tilde \Gamma[0,V]$.
%
\subsection{Physical equivalence of the connected background
functional} \label{sec:pheq}
The physical equivalence of the connected background
functional $W_{\rm bg}$ has been proved by Becchi and Collina \cite{becchi}.
Here we provide a shorter proof.

We comment on the relationship between 
$W_{\rm bg}$ in eq.(\ref{ab9cpp}) and
$W$ in eq.(\ref{ab6}).
From the definition of $\tilde W$ in eq.(\ref{ab8})
and by using eq.(\ref{ab14}) we obtain:
\begin{eqnarray}
\back W[\bar J[0,V,\beta],V,\beta,\dots] &=& \tilde\G[0,V,\beta,\dots]
+ \int d^4 x \, \bar J[0,V,\beta]V 
\nonumber\\
&&+ 
\int d^4x \left(
K[0,V,\beta]~B - \eta[0,V,\beta]~{\bar \theta} - {\bar\eta}[0,V,\beta]~\theta
\right) \, .
\nonumber\\
\label{ab22}
\end{eqnarray}
Notice that $(K,B)$, $(\eta,\bar\theta)$, $(\bar\eta,\theta)$ are
conjugate variables for the functional $\tilde W$ in eq.(\ref{ab14}).
Moreover, since $\tilde A=0$ we get from eq.(\ref{ab11p}):
\begin{eqnarray}\label{star}
A=\frac{\delta W[J,V,\dots]}{\delta J}|_{J=\bar J[0,V,\beta]}=V\ .
\end{eqnarray}
Now we differentiate both sides of eq.(\ref{ab22}) with respect to $V$. 
By using eq.(\ref{star})  we obtain
\begin{eqnarray}\label{star2}
- \frac{\delta \tilde\Gamma[0,V,\beta]}{\delta V} &= &
-\frac{\delta W[J,V,\beta]}{\delta V}|_{J=\bar J[0,V,\beta]} + 
\bar J[0,V,\beta] \, .
\end{eqnarray}
Now, by using eq.(\ref{star}), it follows from eq.(\ref{ab22}) 
by explicit computation that 
\begin{eqnarray}
\frac{\delta^{(n)}W[J,V,\beta,\dots]}
     {\delta \beta_{i_1}(x_1) \dots \delta \beta_{i_n}(x_n)}
|_{J=\bar J[0,V,\beta]} = 
\frac{\delta^{(n)}\tilde\G[0,V,\beta,\dots]}
     {\delta \beta_{i_1}(x_1) \dots \delta \beta_{i_n}(x_n)} \, .
\label{ab23}
\end{eqnarray}
From the properties of the Legendre transform one also gets:
\begin{eqnarray}
\back\frac{\delta^{(n)}\tilde\G[0,V,\beta,\dots]}{\delta \beta_{i_1}(x_1) \dots \delta \beta_{i_n}(x_n)} =
\frac{\delta^{(n)}W_{\rm bg}[J,\beta,\dots]}{\delta \beta_{i_1}(x_1) \dots \delta \beta_{i_n}(x_n)} 
&\quad&
\mbox{for }
J = -\frac{\delta \tilde \G[0,V,\beta,\dots]}{\delta V}
\label{ab24}
\end{eqnarray}
and finally, by combining eqs.(\ref{ab23}) and (\ref{ab24}),
\begin{eqnarray}
\frac{\delta^{(n)}W[J,V,\beta,\dots]}
     {\delta \beta_{i_1}(x_1) \dots \delta \beta_{i_n}(x_n)}
|_{J=\bar J[0,V,\beta]}
 =
\frac{\delta^{(n)}W_{\rm bg}[J,\beta,\dots]}
     {\delta \beta_{i_1}(x_1) \dots \delta \beta_{i_n}(x_n)} \, 
|_{J=\frac{\delta \tilde \G[0,V,\beta,\dots]}{\delta V}} .
\label{ab25}
\end{eqnarray}
We now set $\beta=0$.
 We can always set
$\bar J[0,V,\beta]|_{\beta=0}=0$ (by choosing $V=0$).

By the arguments of Section~\ref{sec:bfm} 
\begin{eqnarray}
\left.\frac{\delta W[J,V]}{\delta V}\right|_{J=0} = 0 
\label{ab27}
\end{eqnarray}
and from eq.(\ref{star2}) this implies
$\frac{\delta \tilde \G}{\delta V}[0,V,\beta]|_{V=0=\beta}=0$.
 Thus from eq.(\ref{ab25})
\begin{eqnarray}
\left.
\frac{\delta ^{(n)}W[0,V,\beta,\dots]}
    {\delta \beta_{i_1}(x_1) \dots \delta \beta_{i_n}(x_n)}
\right|_{\beta=0}
 =
\left.
\frac{\delta ^{(n)}W_{\rm bg}[J,\beta,\dots]}
    {\delta \beta_{i_1}(x_1) \dots \delta \beta_{i_n}(x_n)}
\right|_{J=0=\beta}
 \, ,
\label{ab28}
\end{eqnarray}
for all values of $V$. 
This is the formal proof that on-shell 
physical connected Green functions generated 
by $W$ can be obtained from the connected amplitudes generated by $W_{\rm bg}$.

%
%
\section{Removal of doublets}
\label{sec:doublets}
This section deals with a rather technical but otherwise
important issue connected with the introduction of
the BRST transformation of $V_{a\mu}$. In the present
approach $V_{a\mu}$ forms a ``doublet'' together with $\Omega_{a\mu}$.
\par
The problem we want to discuss can be formulated in various
ways. Roughly speaking one can ask whether the introduction
of the new field $\Omega$ might modify the Slavnov-Taylor identities
thus depriving the theory of the invariance necessary for
the accomplishment of the renormalization procedure and, in the
present case, of the essential tool for the proof of the
background equivalence theorem.
\par
Fortunately one can prove that the anomaly of the
Slavnov-Taylor is not modified in a substantial way
by the introduction of one or more doublets. The assumptions
which guarantee the validity of this results will be
put in evidence during the proof.
The removal of the doublets is a standard procedure~\cite{Zumino} and is
reported here for completeness. Actually our formulation is a pure
algebraic one and therefore no use of power-counting is made.
Our proof has the virtue to emphasize that the removal of doublets 
($V$, $\Omega$) only requires a linear gauge-fixing term such that the full
dependence of the classical action on the antifields $A^*$ is through the
combination:
\begin{eqnarray}
A_{a\mu}^* + \alpha D_\mu(V)_{ab} \bar \theta_b \,.
\end{eqnarray}
In particular no further assumption on the gauge group structure is required.
\par
Let us assume that the ST identities have been reestablished
up to order $n-1$. By the Quantum Action Principle, the $n$-order
ST breaking term $\Delta^{(n)}\equiv{\cal S} (\Gamma)^{(n)}$
is a local functional of the fields and the external sources
and of their derivatives.
Moreover it obeys the Wess-Zumino consistency condition
\begin{eqnarray}
{\cal S}_0 (\Delta^{(n)})=0 \, .
\label{bd1}
\end{eqnarray}
${\cal S}_0$ denotes the linearized classical ST operator given by
\begin{eqnarray}
\back{\cal S}_0 &=& \int d^4x \,
\left ( D_\mu(A)_{ai} \theta_i \frac{\delta }{\delta A_{a\mu}}
-{1 \over 2} f_{aij} \theta_i \theta_j \frac{\delta }{\delta \theta_a}
+ B_a \frac{\delta }{\delta \bar \theta_a} + \Omega_{a\mu}
\frac{\delta }{\delta V_{a\mu}} 
\right.
\nonumber\\&&
\left.
~~~~~~~~~~~~+ \frac{\delta \G^{(0)}}{\delta A_{a\mu}}
\frac{\delta }{\delta A^*_{a\mu}} +
\frac{\delta \G^{(0)}}{\delta \theta_a}\frac{\delta }{\delta \theta^*_a}
\right ) \, .
\nonumber\\
\label{q00}
\end{eqnarray}
If the ST identities are not restored at lower orders, $\Delta^{(n)}$ turns
out to be a non-local functional and the Wess-Zumino consistency condition
is modified with respect to eq.(\ref{bd1}) by non-local
 contributions~\cite{pq}.
\par
In order to study the dependence of the cohomology of
${\cal S}_0$ on the doublets $(V_{a\mu},\Omega_{a\mu})$
we perform the following change of variables.
We replace the antifields
$A_{a\mu}^*$ with
\begin{eqnarray}
\hat A_{a\mu}^* \equiv A_{a\mu}^* + \alpha D_\mu(V)_{ab} \bar \theta_b \,
\label{q1}
\end{eqnarray}
while we keep all other fields and antifields unchanged.
This transformation is invertible. We introduce the operator ${\cal T}$
such that
$$
\hat X(A^*_{a\mu},\phi) \equiv
{\cal T}X(A^*_{a\mu},\phi) =
  X(A_{a\mu}^* + \alpha D_\mu(V)_{ab} \bar \theta_b,\phi) =
  X(\hat A^*_{a\mu},\phi)\, .
$$

Then we define $\hat {\cal S}_0 = {\cal T}{\cal S}_0{\cal T}^{-1}$,
and we find
\begin{eqnarray}
\back\hat {\cal S}_0 &=& \int d^4x \,
\left ( D_\mu(A)_{ai} \theta_i \frac{\delta }{\delta A_{a\mu}}
-{1 \over 2} f_{aij} \theta_i \theta_j \frac{\delta }{\delta \theta_a}
+ B_a \frac{\delta }{\delta \bar \theta_a} + \Omega_{a\mu}
\frac{\delta }{\delta V_{a\mu}} \right . \nonumber \\
&& \left . + \left [ \frac{\delta \G^{(0)}}{\delta A_{a\mu}}
+\alpha \partial_\mu B_a + \alpha f_{aij} \Omega_{i\mu} \bar \theta_j
+\alpha f_{aij} V_{i\mu} B_j
\right ]
\frac{\delta }{\delta \hat A^*_{a\mu}} +
\frac{\delta \G^{(0)}}{\delta \theta_a}\frac{\delta }{\delta \theta^*_a}
\right ) \, .
\label{q2}
\end{eqnarray}
Notice that eq.(\ref{bd1}) tells us that
\begin{equation}\label{bd1bis}
\hat {\cal S}_0 {\cal T}\Delta^{(n)}= {\cal T} {\cal S}_0 \Delta^{(n)} = 0\,
.
\end{equation}
We introduce the operator
\begin{eqnarray}
{\cal K}= \int_0^1 dt~ V_{\mu
a}~\lambda_t~\frac{\delta}{\delta\Omega_{a\mu}}
\label{bd4}
\end{eqnarray}
where the action of $\lambda_t$
 on a generic functional $X(V,\Omega,\hat A^*,\phi)$ is given by
\begin{eqnarray}
\lambda_t \left(X(V,\Omega,\hat A^*,\phi)\right) =
 X(tV,t\Omega,\hat A^*,\phi)
\label{bd5}
\end{eqnarray}
being $\phi$ any other field or source~\footnote{Notice that
$ X(t V,t \Omega,\hat A^*,\phi) \neq
 \hat X(t V,t \Omega,A^*,\phi)$, i.e.
$[\lambda_t,{\cal T}]\neq 0$.}.
By explicit computation one verifies that
\begin{eqnarray}\label{SK_KS}
\back\{\hat {\cal S}_0 ,{\cal K}\} &=& 
\int_0^1 dt\ \left(V\lambda_t \frac{\delta }{\delta V} + \Omega\lambda_t\frac{\delta }{\delta \Omega}\right)+
\nonumber\\&&
\int_0^1 dt\ V_b\lambda_t
        \left(\frac{\delta ^2\G^{(0)}}{\delta A_a\delta\Omega_b}+\alpha f_{abc}\bar\theta_c\right)
        \frac{\delta }{\delta \hat A^*_a}
+
\int_0^1 dt\ V_b\lambda_t
        \frac{\delta ^2\G^{(0)}}{\delta \theta_a\delta\Omega_b}
        \frac{\delta }{\delta \theta^*_a}\,,
\end{eqnarray}
and due to the fact that 
\begin{eqnarray}
\frac{\delta ^2\G^{(0)}}{\delta \theta_a\delta\Omega_b}=0 &\mbox{and }&
\frac{\delta ^2\G^{(0)}}{\delta A_a\delta\Omega_b}=-\alpha f_{abc}\bar\theta_c
\end{eqnarray}
we obtain
\begin{eqnarray}
\{\hat {\cal S}_0 ,{\cal K}\} = 
\int_0^1 dt\ \left(V\lambda_t \frac{\delta }{\delta V} + \Omega\lambda_t\frac{\delta }{\delta \Omega}\right)
\,.
\end{eqnarray}
When applied to the functional $\Delta^{(n)}(V,\Omega,\hat A^*, \phi)$
the last equation gives:%
\footnote{Notice that
$X(0,0,\hat A^*,\phi) \neq \hat X(0,0,A^*,\phi)$ .
}
\begin{eqnarray}
\{\hat {\cal S}_0 ,{\cal K}\}
\Delta^{(n)}(V,\Omega,\hat A^*, \phi) &=&
   \Delta^{(n)}(V,\Omega,\hat A^*, \phi)
  - \Delta^{(n)}(0,0,\hat A^*,\phi) \, .
\label{bd5a}
\end{eqnarray}
Then
\begin{eqnarray}
\Delta^{(n)}(V,\Omega,\hat A^*, \phi) & = &
 \Delta^{(n)}(0,0,\hat A^*, \phi) +
 \hat {\cal S}_0 {\cal K}\Delta^{(n)}(V,\Omega,\hat A^*, \phi)
\nonumber\\&&
~~~~~~~~~~~~~ +{\cal K}\hat {\cal S}_0 \Delta^{(n)}(V,\Omega,\hat A^*, \phi)
\label{bd6}
\end{eqnarray}
The last term
 is vanishing according to eq.(\ref{bd1bis}).

By applying ${\cal T}^{-1}$ to eq.(\ref{bd6}) we obtain:
\begin{eqnarray}
\Delta^{(n)}(V,\Omega,A^*, \phi) & = &
 \Delta^{(n)}(0,0,A_{a\mu}^* + \alpha D_\mu(V)_{ab} \bar \theta_b, \phi) +
 {\cal S}_0 [{\cal T}^{-1}{\cal K}{\cal T} \Delta^{(n)}]
\end{eqnarray}
The dependence on $\Omega_{a\mu}$ is confined to the cohomologically
trivial term 
$${\cal S}_0 [{\cal T}^{-1}{\cal K}{\cal T}\Delta^{(n)}]\,,$$
and $\Delta^{(n)}$ depends non-trivially on $V_{a\mu}$ only through
the dependence on $\hat A^*_{a\mu}$.

By the same technique we show that the non-trivial dependence
of $\Delta^{(n)}$ on $\bar \theta$ is only through
the dependence on $\hat A^*_{a\mu}$. Moreover, there is no
non-trivial dependence on $B_a$.

For this purpose we introduce the new homotopy
\begin{eqnarray}
{\cal K}_B= \int_0^1 dt~ \bar
\theta_{a}~\lambda_t~\frac{\delta}{\delta B_{a}}
\label{bd4bis}
\end{eqnarray}
where now $\lambda_t$ refers to the doublet $(\bar \theta_a, B_a)$:
\begin{eqnarray}
\lambda_t X(\bar \theta,B,\hat A^*,\phi) =  X(t \bar \theta,t B,\hat
A^*,\phi)
\label{bd5bis}
\end{eqnarray}
being $\phi$ any other field or source.
A computation analogous to eq.(\ref{SK_KS}) now yields:
\begin{eqnarray}\label{SKB_KBS}
\back\{\hat {\cal S}_0 ,{\cal K}_B\} &=& 
\int_0^1 dt\ \left(\bar\theta\lambda_t \frac{\delta }{\delta \bar\theta} 
                 + B\lambda_t\frac{\delta }{\delta B}\right)
+\nonumber\\&&
\int_0^1 dt\ \bar\theta_b\lambda_t
        \left(\frac{\delta ^2\G^{(0)}}{\delta A_a\delta B_b}+\alpha f_{abc}V_c\right)
        \frac{\delta }{\delta \hat A^*_a}
+
\int_0^1 dt\ \bar\theta_b\lambda_t
        \frac{\delta ^2\G^{(0)}}{\delta \theta_a\delta B_b}
        \frac{\delta }{\delta \theta^*_a}\,,
\end{eqnarray}
and due to the fact that 
\begin{eqnarray}
\frac{\delta ^2\G^{(0)}}{\delta \theta_a\delta B_b}=0 &\mbox{and }&
\frac{\delta ^2\G^{(0)}}{\delta A_a\delta B_b}=-\alpha f_{abc}V_c
\end{eqnarray}
we obtain
\begin{eqnarray}
\{\hat {\cal S}_0 ,{\cal K}_B\} = 
\int_0^1 dt\ \left(\bar\theta\lambda_t \frac{\delta }{\delta \bar\theta}
                 + B\lambda_t\frac{\delta }{\delta B}\right)
\,.
\end{eqnarray}
When applied to the functional $\Delta^{(n)}(V,\Omega,\hat A^*, \phi)$
the last equation gives:
\begin{eqnarray}
\{\hat {\cal S}_0 ,{\cal K}_B\}
\Delta^{(n)}(V,\Omega,\hat A^*, \phi) &=&
        \Delta^{(n)}(\bar\theta ,B,\hat A^*, \phi)
   -    \Delta^{(n)}(0,0,\hat A^*,\phi) \, ,
\label{bd5abis}
\end{eqnarray}
and we get
\begin{eqnarray}
\Delta^{(n)}(\bar \theta,B,\hat A^*, \phi) &=&
\Delta^{(n)}(0,0,\hat A^*, \phi) + \hat {\cal S}_0 {\cal K}_B\Delta^{(n)}
+{\cal K}_B\hat {\cal S}_0 \Delta^{(n)}\nonumber\\
&=&
\Delta^{(n)}(0,0,\hat A^*, \phi) + \hat {\cal S}_0 {\cal K}_B\Delta^{(n)}
 \, ,
\label{bd6bis}
\end{eqnarray}
which gives the announced result.

Notice that the same dependence on $B,\bar\theta$ and $A^*$ for
$\Delta^{(n)}$
is obtained if one requires
that the quantum effective action $\G$ fulfills the ghost
equations:
\begin{eqnarray}
G_a(\G) \equiv \left(\frac{\delta }{\delta \bar\theta_a}- \alpha D_{\mu}(V)_{ab}
\frac{\delta }{\delta A^*_{b\mu}}\right)\G =
(\alpha' - \alpha) (\partial^\mu \Omega_{a\mu})
-\alpha f_{abc} A^\mu_b \Omega_{c\mu} \, .
\label{qg}
\end{eqnarray}
Eq.(\ref{qg}) implies for $n\geq 1$ that
\begin{eqnarray}\label{ghost}
G_a (\G^{(n)}) = 0 \, .
\end{eqnarray}
Moreover, by explicit computation one gets:
\begin{eqnarray}
\back \left\{ {\cal S}_0, G_a \right \} &=&
- \alpha f_{aij} \Omega^\mu_i \frac{\delta }{\delta A_{j\mu}^*}
+ \int d^4x \,\left(  \left ( G_a \G^{(0)}_{A^\mu_i} \right )
\frac{\delta }{\delta A_{i\mu}^*} +
\left ( G_a \G^{(0)}_{\theta_i} \right ) \frac{\delta }{\delta \theta^*_i}
\right) \nonumber \\
& = &
- \alpha f_{aij} \Omega^\mu_i \frac{\delta }{\delta A_{j\mu}^*}
 + \int d^4x \, \left( \frac{\delta }{\delta A^\mu_i} \left ( G_a \G^{(0)} \right )
 \frac{\delta }{\delta A_i^{* \mu}}
- \frac{\delta }{\delta \theta_i} \left ( G_a \G^{(0)} \right ) \frac{\delta }{\delta \theta^*_i}\right)
 \nonumber \\& &
+ \int d^4x \,\left(  \left [ G_a , \frac{\delta }{\delta A^\mu_i} \right ]
\G^{(0)} \frac{\delta }{\delta A_{i\mu}^*}
+ \left \{  G_a , \frac{\delta }{\delta \theta_i} \right \} \G^{(0)}
\frac{\delta }{\delta \theta^*_i} \right)
 \nonumber \\
& = &  - \alpha f_{aij} \Omega^\mu_i \frac{\delta }{\delta A_{j\mu}^*}
+  \alpha f_{aij} \Omega^\mu_i \frac{\delta }{\delta A_{j\mu}^*} = 0 \, ,
\label{qg1}
\end{eqnarray}
since
\begin{eqnarray}
 \left [ G_a , \frac{\delta }{\delta A^\mu_i} \right ] =
 \left \{  G_a , \frac{\delta }{\delta \theta_i} \right \} = 0 \, .
\label{qg2}
\end{eqnarray}
Then $G_a(\G^{(n)})=0$ implies $G_a(\Delta^{(n)})=0$.
Notice that the converse is not true: $\Delta^{(n)}$ is not modified if one
adds
to $\Gamma^{(n)}$ an action-like term of the form
$$\Xi^{(n)} \equiv {\cal S}_0(\int d^4x \, \bar \theta_a H^a) \, ,$$
where $H^a$ is a FP-charge $0$
Lorentz-invariant polynomial in the fields and the external sources of the
model
with dimension $\leq 2$.
However, $\Xi^{(n)}$ could spoil the ghost equations at order $n$ (take for
example
$H_a = \partial_\mu V^{\mu}_a$).

One is thus led to study the following consistency condition
\begin{eqnarray}
S_0 \Delta^{(n)}(A^*_{a\mu} + \alpha D_{\mu}(V)_{ab} \bar \theta_b,
A_{a\mu},\theta_a, \theta^*_a) = 0 \, .
\label{bd8}
\end{eqnarray}
This can be explicitly recast as:
\begin{eqnarray}
0&=&\int d^4x \,
\left ( D_\mu(A)_{ai} \theta_i \frac{\delta }{\delta A_{a\mu}}
-{1 \over 2} f_{aij} \theta_i \theta_j \frac{\delta }{\delta \theta_a}
+ B_a \frac{\delta }{\delta \bar \theta_a} + \Omega_{a\mu}
\frac{\delta }{\delta V_{a\mu}} \right . \nonumber \\
&& \quad \quad \quad \left . + \frac{\delta \G^{(0)}}{\delta A_{a\mu}}
\frac{\delta }{\delta A^*_{a\mu}} +
\frac{\delta \G^{(0)}}{\delta \theta_a}\frac{\delta }{\delta \theta^*_a}
\right) \Delta^{(n)}(A^*_{a\mu} + \alpha D_\mu(V)_{ab} \bar \theta_b)
\nonumber\\
&=&\int d^4x \,
\left ( \left [ \frac{\delta \G^{(0)}}{\delta A_{a\mu}}
+\alpha \partial_\mu B_a + \alpha f_{aij} \Omega_{i\mu} \bar \theta_j
+\alpha f_{aij} V_{i\mu} B_j
\right ]
\frac{\delta }{\delta \hat A^*_{a\mu}}\right . \nonumber \\
&& \quad\quad\quad +\left .
 D_\mu(A)_{ai} \theta_i \frac{\delta }{\delta A_{a\mu}}
-{1 \over 2} f_{aij} \theta_i \theta_j \frac{\delta }{\delta \theta_a}
+
\frac{\delta \G^{(0)}}{\delta \theta_a}\frac{\delta }{\delta \theta^*_a}
\right) \Delta^{(n)}(\hat A^*_{a\mu})
\\
&=&\int d^4x \,
\left ( \left [ \frac{\delta }{\delta A_{a\mu}}\G^{(0)}(\hat A^*_{a\mu},
A_{a\mu},\theta_a, \theta^*_a)_{\big|_{B=0,\Omega=0}}
\right ]
\frac{\delta }{\delta \hat A^*_{a\mu}}\right . \nonumber \\
&& \quad\quad\quad +\left .
D_\mu(A)_{ai} \theta_i \frac{\delta }{\delta A_{a\mu}}
-{1 \over 2} f_{aij} \theta_i \theta_j \frac{\delta }{\delta \theta_a}
+ \frac{\delta \G^{(0)}}{\delta \theta_a}\frac{\delta }{\delta \theta^*_a}
\right) \Delta^{(n)}(\hat A^*_{a\mu})
 \, .\label{solution}
\end{eqnarray}
In the second line we have used the fact that the dependence of
$\Delta^{(n)}$ on $V_{a\mu}, \bar \theta_a$ is only
through $\hat A_{a\mu}$, while
in the last line we have used the fact that the term in the square bracket
is actually independent of $B_a$ and $\Omega_a$.
Also notice that, as a consequence of eq.(\ref{qg}),
$\G^{(0)}$ evaluated at $B=0,\Omega=0$ depends on $V_{a\mu}, \bar \theta_a$
through the antifield $\hat A^*_{a\mu}$ only.

The solution $\Delta^{(n)}(\hat A^*_{a\mu}, A_{a\mu},\theta_a, \theta^*_a)$
to eq.(\ref{solution}), evaluated at
$$\hat A^*_{a\mu} = A^*_{a\mu} + \alpha D_\mu(V=0)_{ab} \bar \theta_b\,,$$
is the solution to the Wess-Zumino consistency condition of the original
 theory without the background fields~\cite{piguet}.
We obtain the anomalous functional for the theory where $V\neq0$ by
evaluating $\Delta^{(n)}(\hat A^*_{a\mu}, A_{a\mu},\theta_a, \theta^*_a)$
at $\hat A^*_{a\mu} = A^*_{a\mu} + \alpha D_\mu(V)_{ab} \bar \theta_b$.

Note that if one considers the solution to eq.(\ref{solution}) belonging
to a subspace of the local functionals in
 $(\hat A^*_{a\mu}, A_{a\mu},\theta_a, \theta^*_a)$ of a given dimension,
one can revert to the anomalous functional for the original theory depending
 on $(A^*_{a\mu},V,\bar \theta, A_{a\mu},\theta_a, \theta^*_a)$ of the same
dimension by applying the transformation in eq.(\ref{q1}), 
since this transformation preserves power-counting.
This means that in this case the algebraic procedure is consistent with
the use of power-counting arguments.

\section{Conclusions}
\label{sec:conclusions}
The background field method quantization turns out to be
a very powerful tool in deriving physical predictions
of gauge theories.

In the background field gauge both the connected amplitude functional and
 1PI vertex function satisfy STI associated to the BRST invariance
and Ward identities associated to the background gauge transformations.
This can be used to prove the validity of the formal change of
 variables on the gauge fields and proves the independence of the
 physical amplitude from the background field.
The whole procedure remains valid even if one introduces an extra
 gauge fixing term
($
 -\alpha' s\, \left[{\bar \theta}_a\partial^\mu
V_{a\mu}\right]
$)
 right at the beginning.

Part of the background effective action for $\tilde A\neq0$ has to be
computed if one wants to go beyond the 1-loop approximation.
The renormalization of these amplitudes must be performed by requiring
the validity of the ST identities \cite{Stora}.
On the other side the ST identities have
 been the essential tool in the proof of the background equivalence theorem
(see Ref.~\cite{abbott2} and Section~\ref{sec:how}).

The Legendre transform for the background effective action
 can be constructed only in presence of the extra gauge fixing term
 and it is shown to provide the same physical amplitude as the original
 one.
However it can not be in general associated to a field theory since
there is no match between the Feynman rules for vertices
inside the 1PI amplitudes and
the vertices connected by the linking propagators.

The present approach has the virtue to allow a
complete proof of the background equivalence theorem
for all physical amplitudes, including
 any expectation value of quasi-local
observable operator (therefore BRST invariant objects) and in particular
the physical S-matrix elements of any gauge theory.

We finally would like to remark that
the present approach further clarifies the r\^ole of the
gauge-fixing condition in perturbative quantum field theory.
\subsection*{Acknowledgments}
We acknowledge a partial financial support by MURST.

\end{document}